# Histopathological imaging features- versus molecular measurements-based cancer prognosis modeling

Sanguo Zhang[1], Yu Fan[1,3], Tingyan Zhong[2,3] and Shuangge Ma[3,*]

[1]School of Mathematics Sciences, University of Chinese Academy of Sciences, Beijing 100049, China

[2]SJTU-Yale Joint Center for Biostatistics, Department of Bioinformatics and Biostatistics, School of Life Sciences and Biotechnology, Shanghai Jiao Tong University

[3]Department of Biostatistics, Yale School of Public Health, New Haven, CT 06520, USA

[*]Correspondence: Shuangge.ma@yale.edu; Tel.: 01-203-785-3119

**Abstract.** For most if not all cancers, prognosis is of significant importance, and extensive modeling research has been conducted. With the genetic nature of cancer, in the past two decades, multiple types of molecular data (such as gene expressions and DNA mutations) have been explored. More recently, histopathological imaging data, which is routinely collected in biopsy, has been shown as informative for modeling prognosis. In this study, using the TCGA LUAD and LUSC data as a showcase, we examine and compare modeling lung cancer overall survival using gene expressions versus histopathological imaging features. High-dimensional regularization methods are adopted for estimation and selection. Our analysis shows that gene expressions have slightly better prognostic performance. In addition, most of the gene expressions are found to be weakly correlated imaging features. It is expected that this study can provide some insight into utilizing the two types of important data in cancer prognosis modeling and into lung cancer overall survival.

# 1. Introduction

For most if not all cancers, various prognosis outcomes, such as overall survival, progression free survival, and time to metastasis, are of essential importance. Accordingly, extensive modeling research has been conducted. In "classic" prognosis studies, low-dimensional demographic, clinical, and environmental risk factors are analyzed, and "standard" regression-based techniques (such as Cox model) are usually sufficient. Despite some successes, it has been well recognized that the complexity of cancer prognosis demands additional data and more sophisticated modeling.

Cancer is a genetic disease. In the past two decades, with the fast development of high-throughput sequencing techniques, molecular data have been extensively collected in cancer studies, and accordingly, molecular data-based prognosis modeling has been accumulating. For example, an investigation of miRNA expression in 104 pairs of primary lung cancers and corresponding noncancerous lung tissues found that high hsa-mir-155 and low hsa-let-7a-2 expressions are correlated with poor survival of lung cancer. The signatures were cross validated using an independent set of adenocarcinomas [1]. Since then, hsa-mir-155 over expression has been reported in thyroid carcinoma, breast cancer, colon cancer, and cervical cancer, indicating its potential for serving as a biomarker for tumor detection and evaluation of prognosis outcome [2]. As another example, the study of genome-wide expression of 100 Non–Small-Cell lung cancer (NSCLC) FFPE samples identified a signature composed of 59 genes, which was strongly associated with prognosis for stage I lung cancer patients. This signature was later proven to be robust for clinical usage [3]. Molecular data are high-dimensional and contain substantial "noises", that is, the majority of measurements are not associated with prognosis. To effectively remove noises, identify relevant effects, and build reliable models using "signals" only, a myriad of high-dimensional statistical techniques has been developed. A popular family of approaches conducts regularization and applies techniques such as penalization, boosting, Bayesian, and thresholding, which can simultaneously achieve estimation and variable selection. Such techniques have demonstrated statistical, numerical, and empirical successes. We refer to published literature [4-6] for reviews and more extensive discussions. With the accumulation of clinical and experimental data, there is increasing knowledge on the functionality of molecular changes. As such, studies have also been conducted using molecular changes that have "prior information", for example, with evidence of being relevant from multiple studies. In this line of work, multiple gene panels have been developed. For example, Jablons and colleagues aimed at developing a prognostic risk score for patients with completely resected lung adenocarcinomas based on genes previously identified in microarray models of NSCLC prognosis. They suggested narrowing the 61-gene panel down to 4 genes [7]. A drawback of molecular data is that it is not as easy to collect: many patients are still concerned with providing tissues for molecular profiling, and the cost of high-throughput profiling is still not "friendly".

A more recent type of data comes from histopathological imaging. In cancer clinical practice, biopsy is routinely conducted, which generates histopathological images. Such images have been long used for definitive diagnosis and staging [8]. They contain rich information on tumors' "micro" properties and surrounding microenvironment, which play important roles in cancer development. In a handful of recent studies, histopathological imaging features have been used for modeling cancer prognosis (as well as other outcomes and phenotypes) [9,10]. However, such studies are still relatively limited. With the consideration that tumor properties can be affected by molecular changes, there have been studies modeling the relationships between imaging features and molecular changes [11,12]. Considering the cost-effectiveness and routineness of biopsy and histopathological images, imaging-based modeling, if effective, can have great potential. Here we distinguish between histopathological images and radiological images – the latter are generated by CT, PET, and other radiological techniques and inform "macro" properties of tumors such as size, shape, and density.

A common limitation of the existing studies is that information has been scattered. More specifically, studies that analyze both histopathological imaging features and molecular changes using the same data and on the same ground are very limited. With differences in patient characteristics and data generation, processing, and analysis procedures, findings from different studies may not be directly comparable.

The objective of this theoretical study is multi-fold. Specifically, it intends to further demonstrates cancer prognosis modeling using histopathological imaging and molecular data, taking advantage of high-dimensional regularization techniques. More importantly, it provides a direct and fair comparison of modeling using these two types of highly important and popular data. To be comprehensive, we also move on and examine integrating these two types of data for modeling prognosis as well as modeling their relationships. With the analysis of TCGA LUAD and LUSC data, this study may also provide additional insight into lung cancer prognosis.

## 2. Materials

TCGA (The Cancer Genome Atlas) is one of the largest and most comprehensive cancer projects organized by the NCI (National Cancer Institute) and NHGRI (National Human Genome Research Institute). For over thirty different types of cancer, it has published comprehensive phenotypic, demographic, molecular, and imaging data [13]. We choose to analyze TCGA data because of its high quality, comprehensiveness, and public availability. In particular, we analyze data on LUAD (lung adenocarcinoma) and LUSC (lung squamous cell carcinoma), two subtypes of NSCLC. Lung cancer patients in general have poor prognosis, and as such, prognosis modeling can be especially important. For prognosis outcome, we choose overall survival, which has also been analyzed in Radzikowska, *et al*. [14], Collins, *et al*.[15], and quite a few other studies.

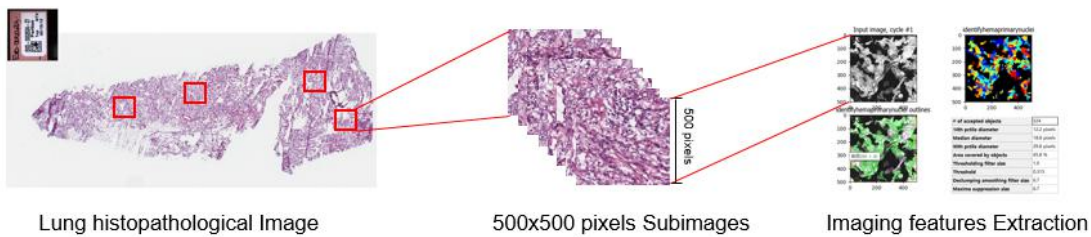

**Figure 1.** Pipeline for extracting imaging features

*2.1 Histopathological imaging data*

Whole-slide histopathology images are downloaded from the TCGA website (http://portal.gdc.cancer.gov) and are in the svs format. These tissue slides are formalin-fixed and paraffin-embedded, and the cell morphology is well-preserved and suitable for image feature recognition. They are captured at 20* or 40* magnification by the Aperio medical scanner. In recent studies, we [12] and others [16,17] have developed and implemented a pipeline for extracting high-dimensional imaging features, which is sketched in Figure 1. Briefly, it includes the following three main steps. First, whole-slide histopathology images are chopped into small subimages of 500*500 pixels, and 20 subimages are randomly selected from the whole slide image. Then, imaging features are extracted using CellProfiler [18], a publicly available software that has been adopted in quite a few recent studies [10,19,20]. In the next step, for each patient, features are averaged. We refer to Zhong, *et al.*[12] and Luo, *et al.*[16] for more detailed discussions on this imaging processing pipeline as well as alternatives. With this processing pipeline, as many as 299 features can be obtained. We note that this is significantly higher than in studies such as Wang, *et al.*[21] and Romo, *et al.*[22], which consider low-dimensional imaging features. Comparatively, high-dimensional features have less lucid interpretations but can contain more information and be more powerful in modeling. As such, they are adopted in this analysis. With the extracted features, we further conduct quality control. In particular, irrelevant features, such as file size and execution information, are removed. We also remove features with severe missingness (>25%) and no or little variation. A total of 221 features are included in downstream analysis.

*2.2 Molecular data*

For molecular data, we analyze gene expressions, which have been considered in many lung cancer prognosis modeling studies [23,24]. Compared to DNA and epigenetic changes, gene expressions are "closer" to phenotypes. With a lack of high-quality protein data, TCGA gene expression data have been extensively analyzed for prognosis and other phenotypes and biomarkers. In TCGA, gene expressions were measured using the Illumina Hiseq2000 RNA Sequencing Version 2 analysis platform and processed and normalized using the RSEM software. More detailed information is available in the literature [25,26]. It is possible to directly conduct whole transcriptome analysis. However, findings may be unreliable when sample sizes are limited. As such, we take a candidate gene approach. In particular, the 61 gene panel developed in Raz, *et al*. [7] is adopted. Matching this panel with gene names in the TCGA data leads to 50 genes for analysis. We acknowledge that there is still a lack of definitive consensus on lung cancer prognosis genes and that there are other lung cancer prognosis gene panels. This particular panel is selected as it has been recently examined in authoritative studies. The proposed analysis can be directly applied to other prognosis panels.

*2.3 Available data*

Beyond imaging and gene expression data, clinical characteristics have also been established as associated with prognosis and included in our analysis. Following published studies and considering data availability, we include sex, age, cancer stage, and tumor size. More specifically, tumor size is defined as longest dimension*shortest dimension, and we combine cancer stages into three levels to avoid small counts. Multiple types of data are combined by matching unique sample IDs. The final LUAD data contains 307 samples. Among them, 106 died, with survival times ranging from 0 to 88.07 months and a median of 20.52 months. There are also 201 censored subjects, with observed times ranging from 0 to 238.11 months and a median of 23.16 months. The final LUSC data contains 334 samples. Among them, 155 died, with survival times ranging from 0.10 to 173.69 months and a median of 18.36 months. There are also 179 censored subjects, with observed times interval ranging from 0.39 to 156.54 months and a median of 23.55 months. For both LUAD and LUSC, data on 221 histopathological imaging features and 50 gene expressions are available. Summary statistics on the clinical characteristics are presented in Table 1.

Table 1. Summary of clinical characteristics.

| | | LUAD (n=307) | | | LUSC (n=334) | | |
|---|---|---|---|---|---|---|---|
| **Sex** | Female | 170 | | Female | 85 | | |
| | Male | 137 | | Male | 249 | | |
| **Age** | | 65.49(sd=9.71) | | | 67.38(sd=8.59) | | |
| **Cancer Stage** | Stage I | 3 | Level_A(164) | Stage I | 1 | Level_A(176) | |
| | Stage IA | 73 | | Stage IA | 60 | | |
| | Stage IB | 88 | | Stage IB | 115 | | |
| | Stage II | 0 | Level_B(77) | Stage II | 1 | Level_B(93) | |
| | Stage IIA | 28 | | Stage IIA | 33 | | |
| | Stage IIB | 49 | | Stage IIB | 59 | | |
| | Stage III | 0 | Level_C(66) | Stage III | 0 | Level_C(65) | |
| | Stage IIIA | 40 | | Stage IIIA | 46 | | |
| | Stage IIIB | 7 | | Stage IIIB | 14 | | |
| | Stage IV | 19 | | Stage IV | 5 | | |
| **Tumor Size** | | 0.467(sd=0.324) | | | 0.470(sd=0.309) | | |

## 3. Analysis techniques

Denote $T$ and $C$ as the event and censoring times, respectively. With right censoring, we observe $(U = \min(T,C), \delta = I(T \leq C))$. Denote $X$ as the $p$-dimensional vector of histopathological imaging features, $Z$ as the $q$-dimensional vector of gene expressions, and $L$ as the $r$-dimensional vector of clinical characteristics. Assume $n$ iid samples.

### 3.1 Associate histopathological imaging features and gene expressions with survival

Here our goal is to conduct various "standard" survival analysis and associate imaging features and/or gene expressions with overall survival, while properly accounting for the effects of clinical characteristics. We comprehensively consider multiple sets of analysis.

First consider the analysis with $X^L = (X', L')'$ as input. Consider the Cox model, under which the hazard function:

$$\lambda(T|X^L) = \lambda_0(T) \exp\left(\boldsymbol{\beta}' X^L\right).$$

Here $\lambda_0(T)$ is the unknown baseline hazard function, and $\boldsymbol{\beta}$ is the vector of unknown regression coefficients. Consider the log partial likelihood function:

$$l(\boldsymbol{\beta}) = \sum_{i=1,\ldots,n} \delta_i \left(\boldsymbol{\beta}' X_i^L - \log\left(\sum_{j=1,\ldots,n} \exp\left(\boldsymbol{\beta}' X_j^L\right) Y_j(U_i)\right)\right),$$

where subscripts $i$ and $j$ correspond to subjects $i$ and $j$, and $Y_j(U_i)$ is the subject $j$'s at risk indicator at time $U_i$. To accommodate the high data dimensionality, and to remove noises and identify relevant effects, we consider the Lasso penalized estimate:

$$\widehat{\boldsymbol{\beta}} = \arg\max\left\{ l(\boldsymbol{\beta}) - \tau \sum_{l=1,\ldots,p} |\beta_l| \right\},$$

where $\tau > 0$ is the data-dependent tuning parameter and chosen using cross-validation, and $\beta_l$ is the $l$th component of $\boldsymbol{\beta}$. Here it is noted that penalization is only imposed on the imaging features. As such, the clinical variables are automatically included, given their established importance in lung cancer prognosis. For a specific imaging feature, a nonzero estimate suggests its association with survival. Literature review suggests that penalization is one of the most popular techniques for accommodating high-dimensional input and feature selection, and Lasso is likely the most popular penalization technique. The adopted "Cox model + Lasso estimation" approach has been examined in multiple published studies [27,28]. In our analysis, it is realized using the R package "glmnet". We note that analysis can also be conducted using other penalties and regularization techniques other than penalization, and that analysis results depend on the adopted technique.

Next we consider the analysis with $Z^L = (Z', L')'$ as input. Analysis can be conducted in the same manner as for imaging features. Denote $\gamma$ as the vector of unknown regression coefficients in the Cox model and $\widehat{\gamma}$ as its Lasso penalized estimate. Note that the baseline hazard functions in this and the above analysis may be different. In this analysis, although the genes have been pre-selected, it is still necessary to apply penalization. In particular, the number of variables, relative to the sample size, is still large. As such, certain regularization is needed in estimation. In addition, to be "cautious", it may still be sensible to examine whether all genes in the panel are associated with survival for the particular TCGA patient cohort (which may differ from those examined in published studies).

In the next set of analysis, we integrate the imaging features and gene expressions using a simple additive approach. In particular, we consider a Cox model with input variable $\left((\hat{\beta}_1,...,\hat{\beta}_p)X, (\hat{\gamma}_1,...,\hat{\gamma}_q)Z, L'\right)'$. Prior to model fitting, we compute the correlation coefficient between $(\hat{\beta}_1,...,\hat{\beta}_p)X$ and $(\hat{\gamma}_1,...,\hat{\gamma}_q)Z$, which can suggest whether the two types of data have overlapping information in modeling survival (after adjusting for the clinical variables). In model fitting, as the dimensionality is low, we do not impose any penalization. This analysis takes a simple additive modeling strategy, which has been developed in the literature [19] and shown as reasonably effective in data integration. It retains the "structure" of imaging effects and that of gene expressions.

For the above three sets of survival analysis, we adopt a random splitting approach to evaluate prediction performance: (a) Randomly split all samples into a training and a testing set with sizes roughly 3:1; (b) Conduct survival analysis as described above using the training set; (c) For subjects in the testing set, compute the predicted risk scores. For example, for the analysis with imaging features, the risk scores are $\hat{\beta}'X^L$. Compute the C-index using the predicted risk scores and testing set (observed time, event indicator). The C-index ranges between 0 and 1, with a larger value indicating better prediction. It is also the time-integrated AUC (Area under the Receiver Operating Characteristic curve). To avoid an extreme split, Steps (a)-(c) are repeated 100 times, and the average C-index is computed to quantify prediction performance. The goal of this analysis is two-fold. The first is to directly compare prognostic performance of the imaging-based model versus that of the gene expression-based. In addition, this analysis also examines whether integrating the two distinct types of measurements can further improve prediction performance.

*3.2 Associate gene expressions with histopathological imaging features*

Tumor properties and surrounding microenvironment are reflected in histopathological images and at least partially regulated by molecular changes. In this sense, imaging features and molecular changes should be interconnected. In recent literature, there have been a handful of studies examining their relationship [12,17]. Such analysis can directly suggest whether the two types of data have overlapping information and its degree. Note that this analysis is "unsupervised", does not involve survival, and can more broadly suggest overlapping information than the correlation analysis above.

Consider the model:

$$\boldsymbol{X} = \eta \boldsymbol{Z} + \boldsymbol{\epsilon},$$

where $\eta$ is the $p \times q$ matrix of regression coefficients, and $\boldsymbol{\epsilon}$ is the $p$-dimensional vector of random errors. Here we model the "downstream" imaging features using the "upstream" gene expressions. Linear regression model is adopted with the consideration that more complex modeling may not be reliable with the limited sample size and high dimensionality of both sides of modeling. For estimating $\eta$, consider:

$$\hat{\eta} = \arg\min \left\{ \sum_{i=1,\dots,n} \|\boldsymbol{X}_i - \eta \boldsymbol{Z}_i\|_2^2 + \tau \sum_{j=1,\dots,q} \|\eta_{j\cdot}\|_2 \right\},$$

where subscript $i$ corresponds to subject $i$, $\tau > 0$ is a data-dependent tuning parameter and chosen using cross-validation, $\eta_{j\cdot}$ is the $j$th row of $\eta$, and $\|\cdot\|_2$ is the $l_2$ norm. Here to accommodate the high data dimensionality and select gene expressions that are relevant for imaging features, we apply the group Lasso penalization.

Similar as above, to more objectively evaluate the relationship, we consider the following approach: (a) Randomly split data into a training and a testing set in the same way as above; (b) Conduct the group Lasso estimation using the training set; (c) For the testing set subjects, predict imaging feature values using gene expressions and the training set estimate. For each imaging feature, compute the correlation coefficient between the predicted and estimated values; (d) To avoid an extreme split, repeat Steps (a)-(c) 100 times, and compute the average correlation values. We note that penalization may introduce shrinkage towards zero. As such, we adopt correlation coefficient as the criterion, which is less affected by shrinkage.

## 4. Results

*4.1 Comparison of modeling using histopathological imaging features or gene expressions*

The first set of analysis regresses survival on the imaging features and clinical characteristics. For the variables included in the final models, their estimated regression coefficients are shown in Tables 2 (LUAD) and 3 (LUSC), respectively. In particular, beyond the clinical characteristics, 7 and 9 imaging features are identified, representing AreaShape, Texture, Granularity, and other characteristics. It has been noted in the literature that, unlike omics and some other types of data, high-dimensional imaging features do not have lucid functional interpretations. As such, we do not further pursue bioinformatics interpretations.

**Table 2.** Analysis of LUAD data: identified imaging features and clinical characteristics associated with overall survival and their estimated coefficients.

| Imaging feature | Coef | Clinical characteristic | Coef |
|---|---|---|---|
| AreaShape_Zernike_6_4 | 0.3697 | Sex | -0.0245 |
| AreaShape_Zernike_8_6 | 0.0426 | Age | 0.0095 |
| AreaShape_Zernike_9_7 | 0.1409 | Tumor_Size | 0.1154 |
| Count_identifytissueregion | 0.1759 | Stage_Level_A | -1.2100 |
| Neighbors_AngleBetweenNeighbors_Adjacent | -0.1033 | Stage_Level_B | -0.2976 |
| Neighbors_FirstClosestObjectNumber_Adjacent | -0.2527 | Stage_Level_C | NA |
| Threshold_WeightedVariance_identifyhemaprimarynuclei | -4.04E-05 | | |

**Table 3.** Analysis of LUSC data: identified imaging features and clinical characteristics associated with overall survival and their estimated coefficients.

| Imaging feature | Coef | Clinical characteristic | Coef |
|---|---|---|---|
| AreaShape_EulerNumber | -0.1575 | Sex | 0.5259 |
| ObjectNumber | -0.2416 | Age | 0.0231 |
| Granularity_12_ImageAfterMath | 0.2382 | Tumor_Size | -0.0369 |
| Threshold_SumOfEntropies_identifytissueregion | 0.1466 | Stage_Level_A | -0.7496 |
| Location_Center_X.1 | -0.0812 | Stage_Level_B | -0.4852 |
| AreaShape_Center_X | -0.0903 | Stage_Level_C | NA |
| AreaShape_Orientation | -0.0985 | | |

| | |
|---|---|
| Neighbors_AngleBetweenNeighbors_Adjacent | 0.1414 |
| Granularity_9_ImageAfterMath | 0.1395 |

In the next set of analysis, we regress survival on gene expressions. The identified gene expressions and clinical characteristics as well as their estimated coefficients are shown in Tables 4 (LUAD) and 5 (LUSC), respectively. Among the identified genes, there are "familiar" discoveries such as PIK3CG [29] and RND3 [30]. In addition, there are also genes that have not yet been well examined in the literature, such as DNMT2 and UQCRC2.

Table 4. Analysis of LUAD data: identified gene expressions and clinical characteristics associated with overall survival and their estimated coefficients.

| Gene expression | Coef | Clinical characteristic | Coef |
|---|---|---|---|
| CCNB1 | 0.0033 | Sex | 0.0011 |
| CTSL | 0.3694 | Age | 0.0173 |
| GLI2 | 0.2555 | Tumor_Size | 0.0640 |
| MFHAS1 | -0.2228 | Stage_Level_A | -1.2460 |
| PIK3CG | -0.3782 | Stage_Level_B | -0.4012 |
| RND3 | 0.1841 | Stage_Level_C | NA |

Table 5. Analysis of LUSC data: identified gene expressions and clinical characteristics associated with overall survival and their estimated coefficients.

| Gene expression | Coef | Clinical characteristic | Coef |
|---|---|---|---|
| IL11 | 0.0526 | Sex | 0.4661 |
| MUC1 | 0.0977 | Age | 0.0309 |
| PIK3CG | 0.0702 | Tumor_Size | -0.4890 |
| PRKCA | 0.1295 | Stage_Level_A | -0.7719 |
| WDHD1 | -0.1404 | Stage_Level_B | -0.6034 |
| | | Stage_Level_C | NA |

When integrating the combined imaging effect with the combined gene expression effect in one Cox model, for the LUAD data, we obtain regression coefficients 0.9842 (imaging feature, p-value=2.12e-6) and 0.4726 (gene expression, p-value=5.36e-9). For the LUSC data, we obtain regression coefficient 0.9709 (imaging feature, p-value=5.55e-9) and 0.8769 (gene expression, p-value=2.04e-3).

In the random-splitting based prediction evaluation, for the LUAD data, the median prediction C-index values are 0.6202 (imaging features), 0.6864 (gene expressions), and 0.6823 (combined). For the LUSC data, the median prediction C-index values are 0.5466 (imaging features), 0.5606 (gene expressions), and 0.5511 (combined). More detailed information, for example on the prediction C-index of each split, is available from the authors.

**Remarks:** In the separate survival analysis with imaging features and gene expressions, relevant effects have been identified. For imaging features, extensive additional research will be needed to fully comprehend the identified variables. We note that this has been noted in the literature [8]. A possible solution is to use low-dimensional features, which may have more lucid interpretations [31]. However, a drawback is that they may contain (much) less information. In the analysis of gene expression data, the "familiarity" of findings may suggest the validity of analysis to a certain extent. However, it is noted that more definitive validation will be needed to confirm the findings. The survival analysis with imaging and gene expression signatures as covariates seems to suggest that the two types of measurements have independent effects. In the random splitting-based evaluation, it is observed that for LUAD, gene expression has moderate predictive performance, and imaging data has moderate/weak predictive performance. For LUSC, both types of measurements have weak predictive performance. For both datasets, gene expression has better performance, which is sensible considering the genetic nature of lung cancer (and other cancers too). Although both LUAD and LUSC are lung cancer subtypes, we observe significantly different results, which can be attributable to the complexity of cancer and suggest that there may not be a definitive conclusion applicable to all cancers. The random splitting evaluation further suggests that integrating the two types of signatures in an additive manner may not further improve prediction, which seems to "contradict" the analysis above. There can be multiple interpretations for this finding. First, the distinction between estimation and prediction should be made – a "good" estimation result may not directly translate into a good prediction. Second, the estimation analysis is repeatedly based on the same data, and there is a risk of over fitting. Third, in the random splitting evaluation, both the training and evaluation are based on fewer observations. An improvement that can be potentially observed with a larger dataset may not be observable with a smaller dataset. It is also noted that penalization and some other sparse approaches have been designed for estimation and may not be ideal for prediction, which may explain the less satisfactory prediction performance observed here.

*4.2 Association of gene expressions and histopathological imaging features*

We first regress imaging features on gene expressions. Detailed information on the identified gene expressions and their estimated coefficients are provided in the Supplementary Materials. In Figure 2, we show the heatmaps of the estimated coefficients. Briefly, for the LUAD data, in the 50*221 coefficient matrix, a total of 7,735 elements are nonzero. A total of 35 genes, including MKI67, ACSL6, NFX1, and WIF1, are identified as associated with 221 imaging features. For the LUSC data, a total of 6,618 elements are nonzero. A total of 28 genes, including ARAF, BCL7A, NXF1, and TP53, are identified as associated with 221 imaging features.

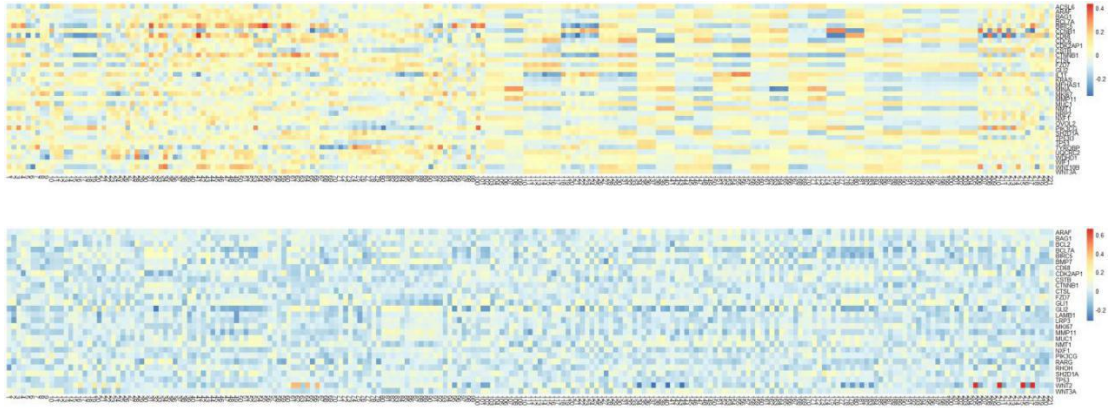

**Figure 2:** Heat map of modeling imaging features using gene expressions. Upper panel: LUAD; Lower panel: LUSC.

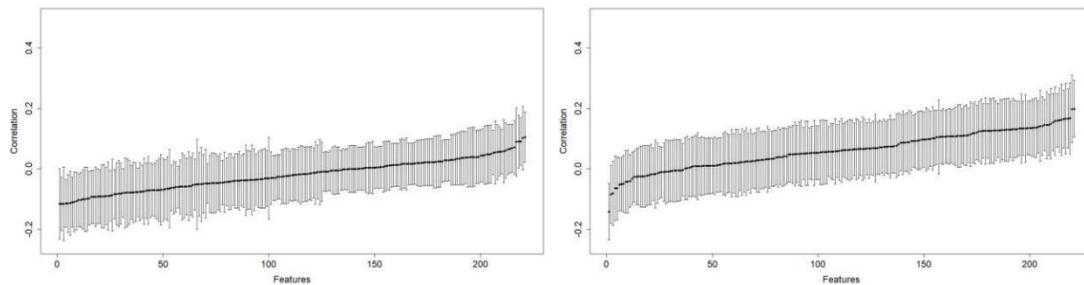

**Figure 3:** Analysis of predicting imaging features using gene expressions: Mean and standard deviation plots of correlation coefficients from 100 random splits. Left: LUAD. Right: LUSC.

The random-splitting based prediction evaluation results are summarized in Figure 3, where we sort performance, from worst to best, across imaging features. More detailed numerical results are provided in the Supplementary Materials.

**Remarks:** The regression analysis suggests that certain gene expressions are connected to imaging features. This observation is sensible considering, as described in Introduction, that properties reflected in imaging features are regulated by molecular changes to a certain extent. On the other hand, the prediction results, as shown in Figure 3, suggest that such associations are mostly weak to moderate. The majority of information in imaging features cannot be readily explained by gene expressions, and this finding differs from that in some published studies [32-34]. It is unclear whether such a difference is attributable to the complexity of cancer, difference in analysis approach, or other factors. More exploration, especially a direct comparison, will be needed.

## 5. Conclusions

Accurately modeling prognosis and other cancer outcomes will remain an important problem for a long time to come. Histopathological imaging data has a great potential. It would be "safe" to predict an increase in imaging feature-based modeling and research that integrate imaging, molecular, clinical, and other data. In this study, we have demonstrated how to analyze/integrate these data under the same ground using advanced statistical techniques. More methodological developments can be built on this article. Our finding on the two lung cancer subtypes has suggested that more sophisticated integration may be needed. The revealed interconnections between imaging and molecular features warrants additional investigation.